\documentclass{PoS}

\title{Measurements and tests on FBK silicon sensors
with an optimized electronic design for a CTA camera}

\ShortTitle{FBK SiPMs for a CTA camera}

\author{
G.~Ambrosi$^{\;(1)}$, M.~Ambrosio$^{\;(2)}$, C.~Aramo$^{\;(2)}$, E.~Bissaldi$^{\; \dagger \,(3,4)}$, M.~Capasso$^{\;(5)}$,  D.~Corti$^{\;(6)}$, A.~de~Angelis$^{\;(6)}$, F.~de~Palma$^{\;(7,8)}$, F.~Ferrarotto$^{\;(9)}$, A.~Ferri$^{\;(10)}$, S.~Garrappa$^{\;(5)}$, N.~Giglietto$^{\;(5,8)}$, \speaker{F.~Giordano\,}$^{\;\; \dagger \;(5,8)}$, A.~Gola$^{\;(10)}$, M.~Ionica$^{\;(1)}$, M.~Iori$^{\;(9,11)}$, F.~Licciulli$^{\;(12)}$, M.~Mariotti$^{\;(6)}$, C.~Marzocca$^{\;(8,12)}$, R.~Paoletti$^{\;(13)}$, C.~Piemonte$^{\;(10)}$, V.~Postolache$^{\;(1)}$, R.~Rando$^{\;(6)}$,  C.~Stella$^{\;(4,14)}$, P.~Vallania$^{\;(15,16)}$, C.~Vigorito$^{\;(15,17)}$ \\
$^{(1)}$INFN-- Sezione di  Perugia;
$^{(2)}$INFN-- Sezione di  Napoli;
$^{(3)}$Dipartimento di Fisica, Universit\`a di Trieste;
$^{(4)}$INFN-- Sezione di  Trieste--Udine;
$^{(5)}$Dipartimento Interateneo di Fisica, Universit\`a e Politecnico di Bari;
$^{(6)}$INFN-- Sezione di  Padova;
$^{(7)}$Universit\`a Telematica Pegaso;
$^{(8)}$INFN-- Sezione di Bari;
$^{(9)}$INFN-- Sezione di  Roma I;
$^{(10)}$FBK-- Trento;
$^{(11)}$Sapienza-- Universit\`a di Roma;
$^{(12)}$Politecnico di Bari;
$^{(13)}$INFN-- Sezione di  Pisa;
$^{(14)}$Dipartimento di Chimica, Fisica e Ambiente, Universit\`a di Udine;
$^{(15)}$INAF-- Osservatorio Astrofisico di Torino;
$^{(16)}$INFN-- Sezione di Torino;
$^{(17)}$Dipartimento di Fisica, Universit\`a di Torino. \\
$\dagger$ {\footnotesize{E-mail:}} {\tt{\footnotesize{Elisabetta.Bissaldi@ts.infn.it, 
Francesco.Giordano@ba.infn.it}}}}
\abstract{In October 2013, the Italian Ministry approved
the funding of a Research \&\ Development (R\&D) study,
within the "Progetto Premiale TElescopi CHErenkov
made in Italy (TECHE)", devoted to
the development of a demonstrator
for a camera for the Cherenkov Telescope Array
(CTA) consortium.
The demonstrator consists of a sensor plane based on the
Silicon Photomultiplier (SiPM) technology and on
an electronics designed for signal sampling.
Preliminary tests on a matrix of sensors produced by
the Fondazione Bruno Kessler (FBK-Trento, Italy)
and on electronic prototypes produced by SITAEL S.p.A.
will be presented. In particular, we used different
designs of the electronics in order to optimize the
output signals in terms of tail cancellation. This
is crucial for applications where a
high background is expected, as for the CTA experiment.}

\FullConference{Science with the New Generation of High Energy Gamma-ray experiments, 10th Workshop - Scineghe2014\\
		04-06 June 2014\\
		Lisbon - Portugal}
\begin{document}
\section{The prototype}
\subsection{The Sensors}
\begin{figure}[ht!]
\centering
\begin{tabular}{cc}
\includegraphics[height=0.55\textwidth,bb=0 0 690 645,clip]{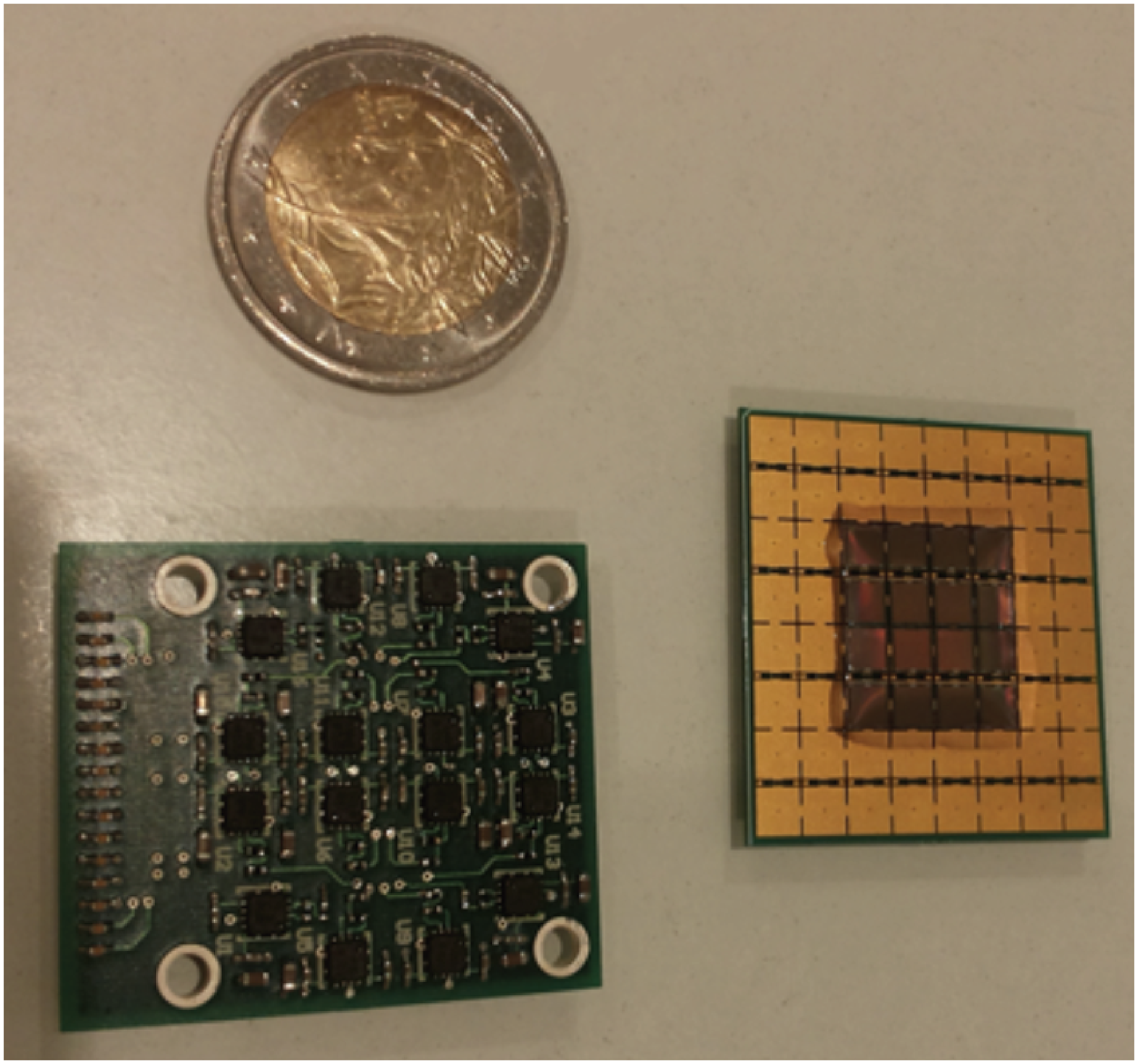} &
\includegraphics[height=0.55\textwidth,bb=0 0 426 732,clip]{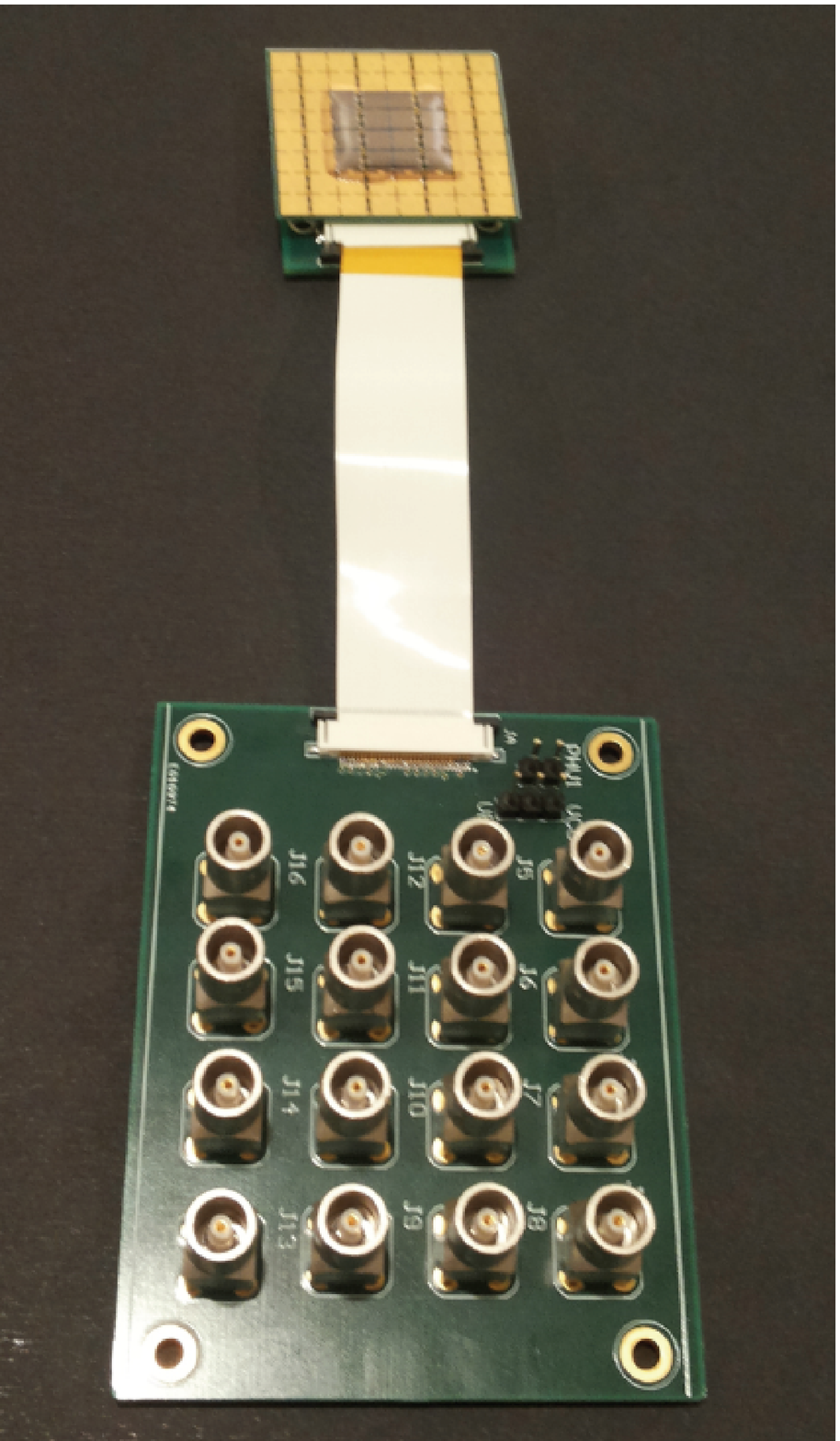}
\end{tabular}
\caption{{\it Left panel}: On the right, we can see
the assembled matrix with the protective layer
of epoxy resin. On the left, the printed circuit board
(PCB) with the
16 AD8000 OPAs. On the top,
Dante's face for size comparison.
{\it Right panel:} Another PCB showing
16 LEMO connectors for the readout.}
\label{Fig_1}
\end{figure}

The devices employed for the realization of the
matrix prototype make use of the FBK \cite{FBK14} Near
Ultra-Violet (NUV) SiPM technology.
The basic structure of these SiPMs
consists of a p$^+-$n junction, whose design
is optimized for the detection
in the blue-NUV region of the electro-magnetic
spectrum \cite{PRO13}.
They show low breakdown voltages
(around 25.6 V) with a slight
temperature dependence of about 24 mV/$^{\rm o}$C.
Moreover, these devices also show a Photon
Detection Efficiency (PDE) of about
24\%\ at 380 nm for 2 V of over-voltage (OV).
In particular, the matrix we tested
consists of 16 SiPMs with 50$\mu$m cells,
for a total equipped area of 12$\times$12 mm$^2$ each
(see Figure \ref{Fig_1}, {\it left panel}).
\subsection{Front End Electronics}
\begin{figure}[ht!]
\centering
\includegraphics[height=0.35\textwidth,bb=0 0 1663 780,clip]{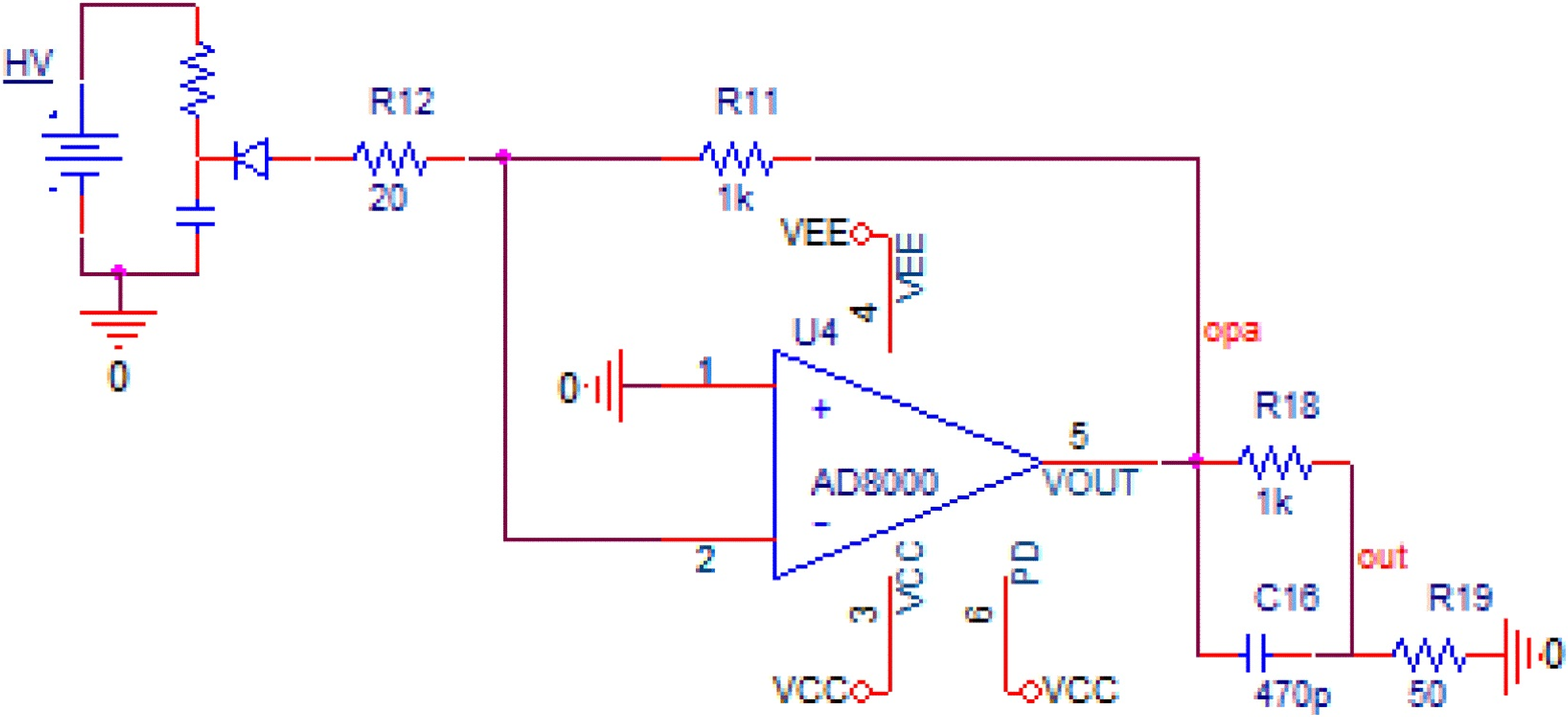} 
\caption{Schematic view of the SiPM and the AD8000 OPA. 
The Pole-Zero cancellation circuit is also drawn.}
\label{Fig_2}
\end{figure}

We designed a preamplifier based on an AD8000 operational
amplifier (OPA) in transimpe-dance configuration.
The design consists of 20 $\Omega$ decoupling resistances
with a 1 k$\Omega$ feedback resistance. The current signal from the SiPM
is fed to the preamplifier in DC mode, as illustrated
in Figure \ref{Fig_2}.

\begin{figure}[hb!]
\centering
\begin{tabular}{cc}
\includegraphics[height=0.24\textheight,bb=0 0 929 720,clip]{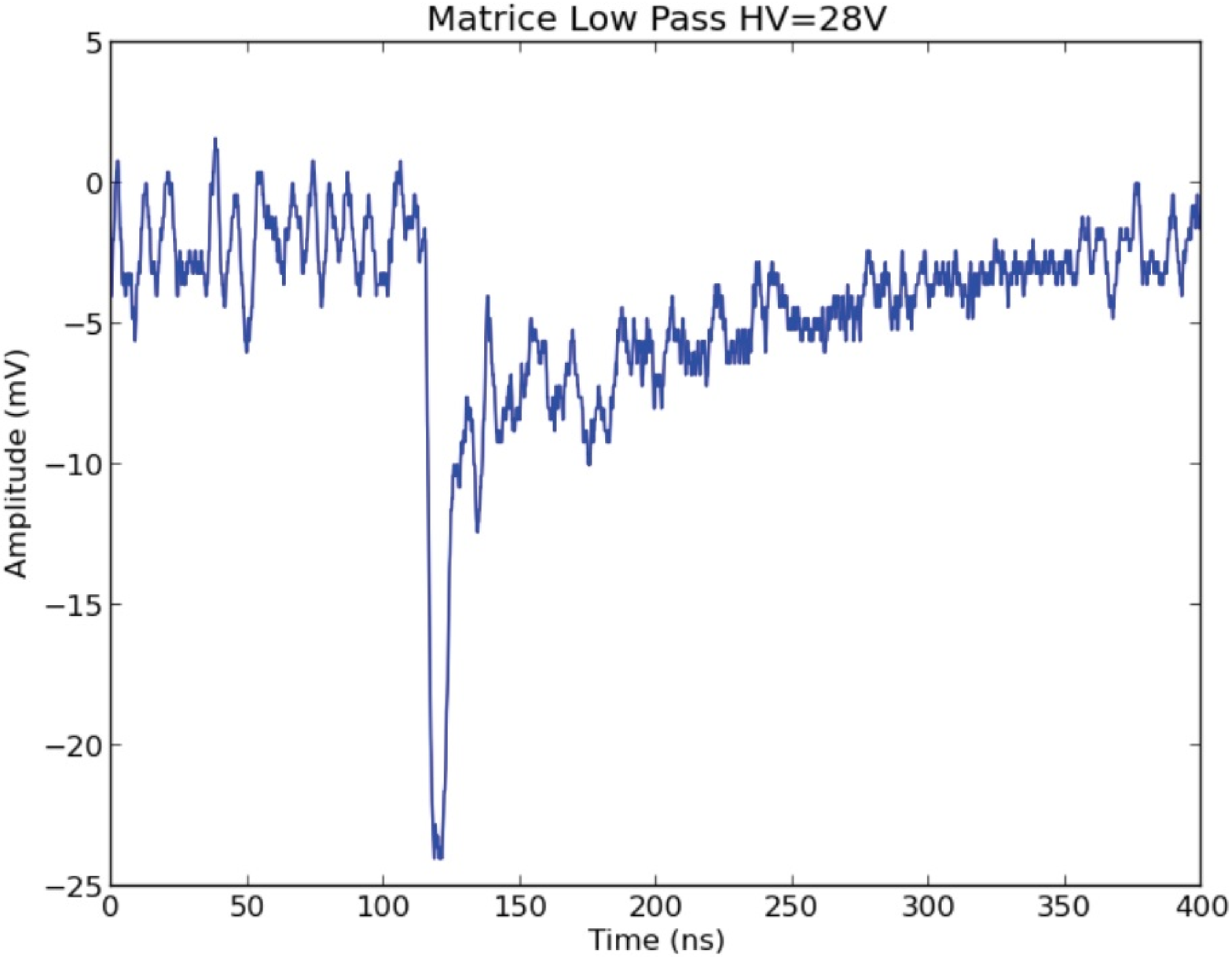} &
\includegraphics[height=0.24\textheight,bb=0 0 930 735,clip]{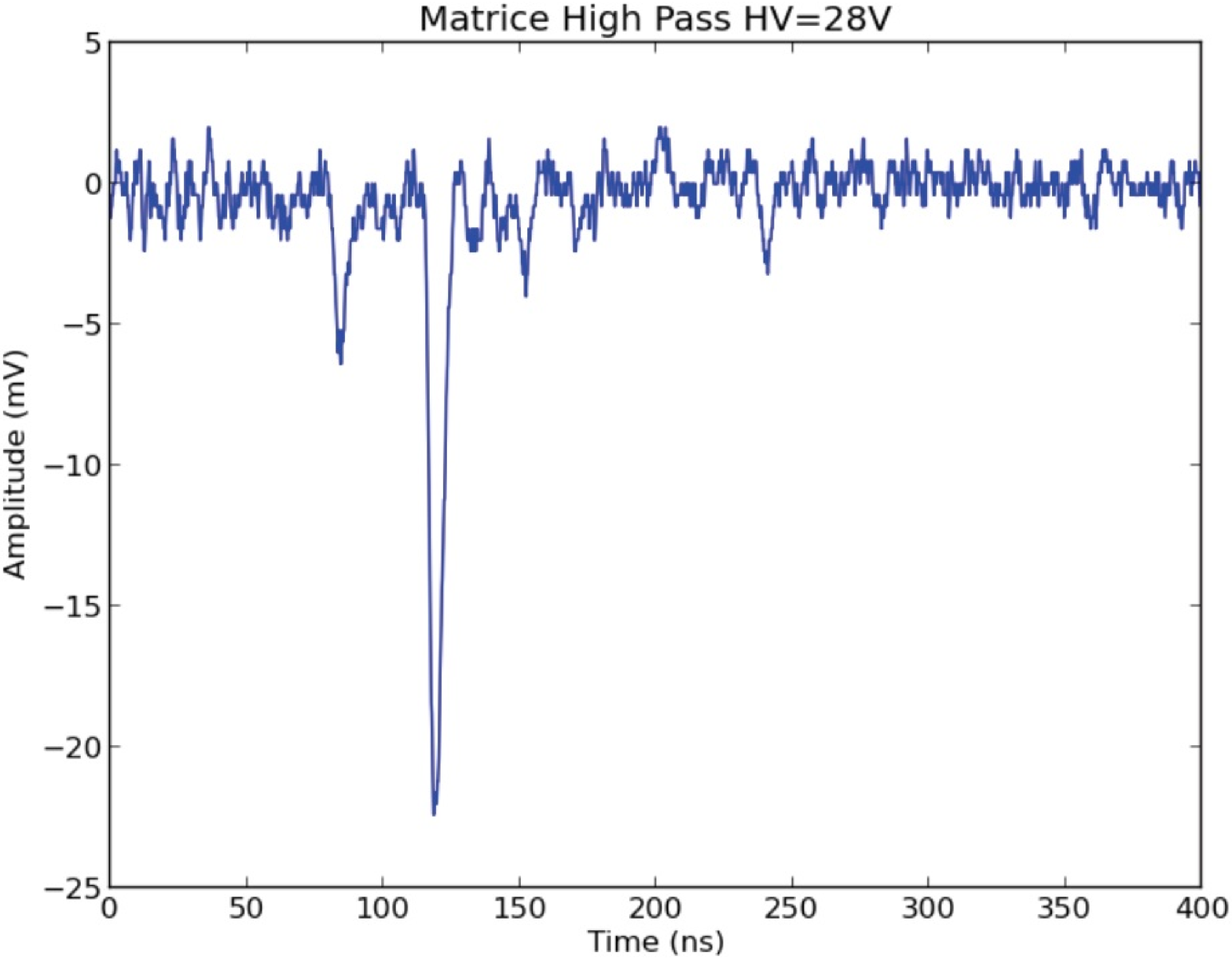}
\end{tabular}
\caption{Current signals from the preamplifier obtained
without ({\it left panel}) and with the Pole-Zero
cancellation circuit ({\it right panel}).}
\label{Fig_3}
\end{figure}

The {\it left panel} of Figure \ref{Fig_3}
shows the current signal
from the preamplifier corresponding
to five photo-electrons (p.e.).
The very fast rise time is due to the development
of the current signal in the SiPM
junction and it is about 100 ps long. Moreover, a
fast decay followed by a very long tail is clearly visible.
This long tail is due to the recovery time
obtained by the product of the quenching resistor times
the single cell capacitance value, and it is about 100 ns
long.
Such a long recovery time can be problematic
in high-background environments, like the one
expected for the CTA experiment.
For this purpose, a Pole-Zero cancellation
circuit \cite{GOL13} was added to the preamplifier
design (see  Figure \ref{Fig_2}). A 470 pF capacitance was placed in
parallel to a 1 k$\Omega$ resistance and closed over
a 50 $\Omega$ load. The signal taken from this Pole-Zero
cancellation network is shown in the {\it right
panel} of Figure \ref{Fig_3}. The absence of the long tail
is clearly visible, and the same amplitude
is almost preserved.
\section{Prototype performance}
\begin{figure}[t!]
\centering
\begin{tabular}{cc}
\includegraphics[height=0.35\textwidth,bb=0 0 914 673,clip]{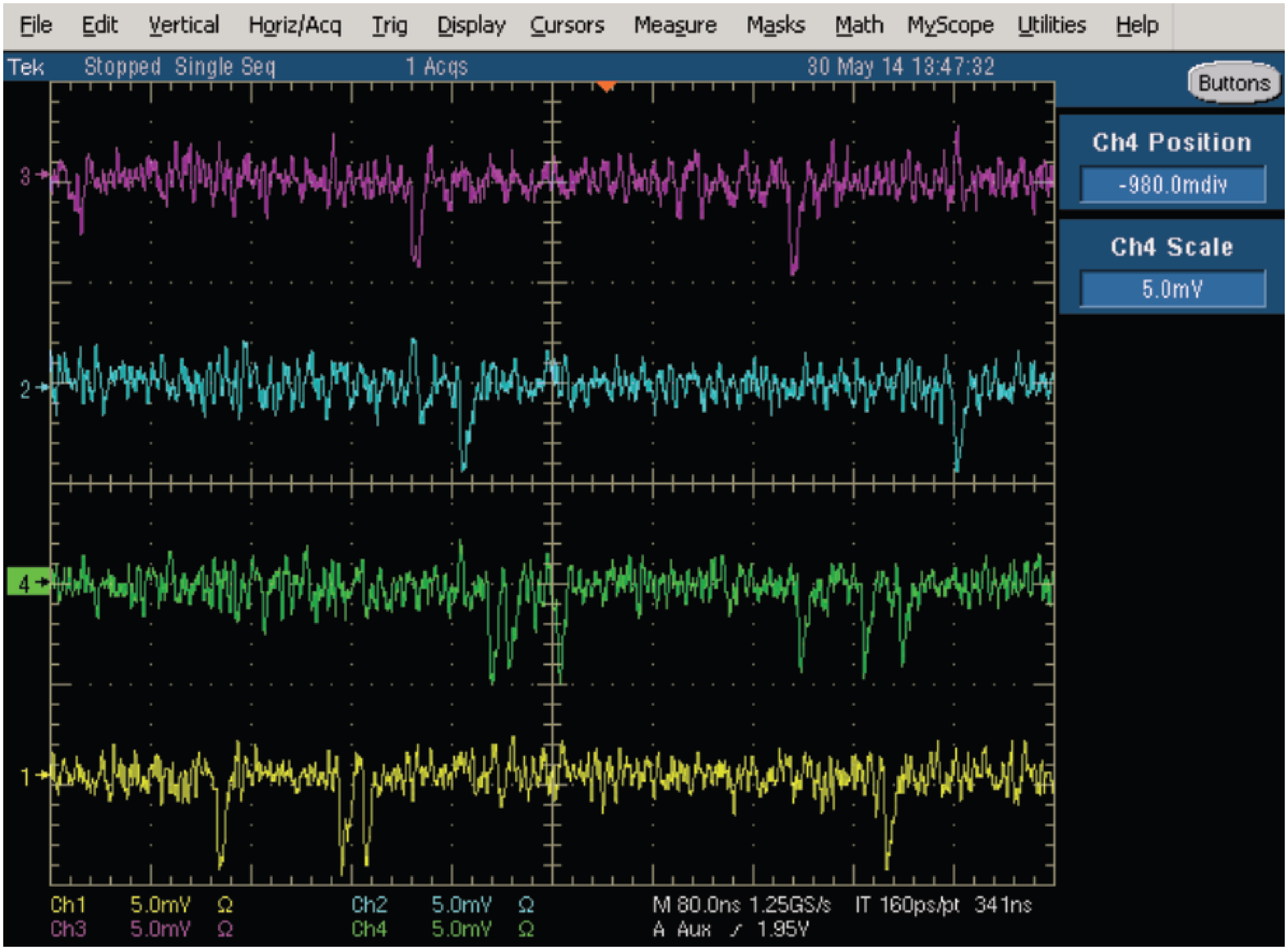} &
\includegraphics[height=0.35\textwidth,bb=0 0 773 580,clip]{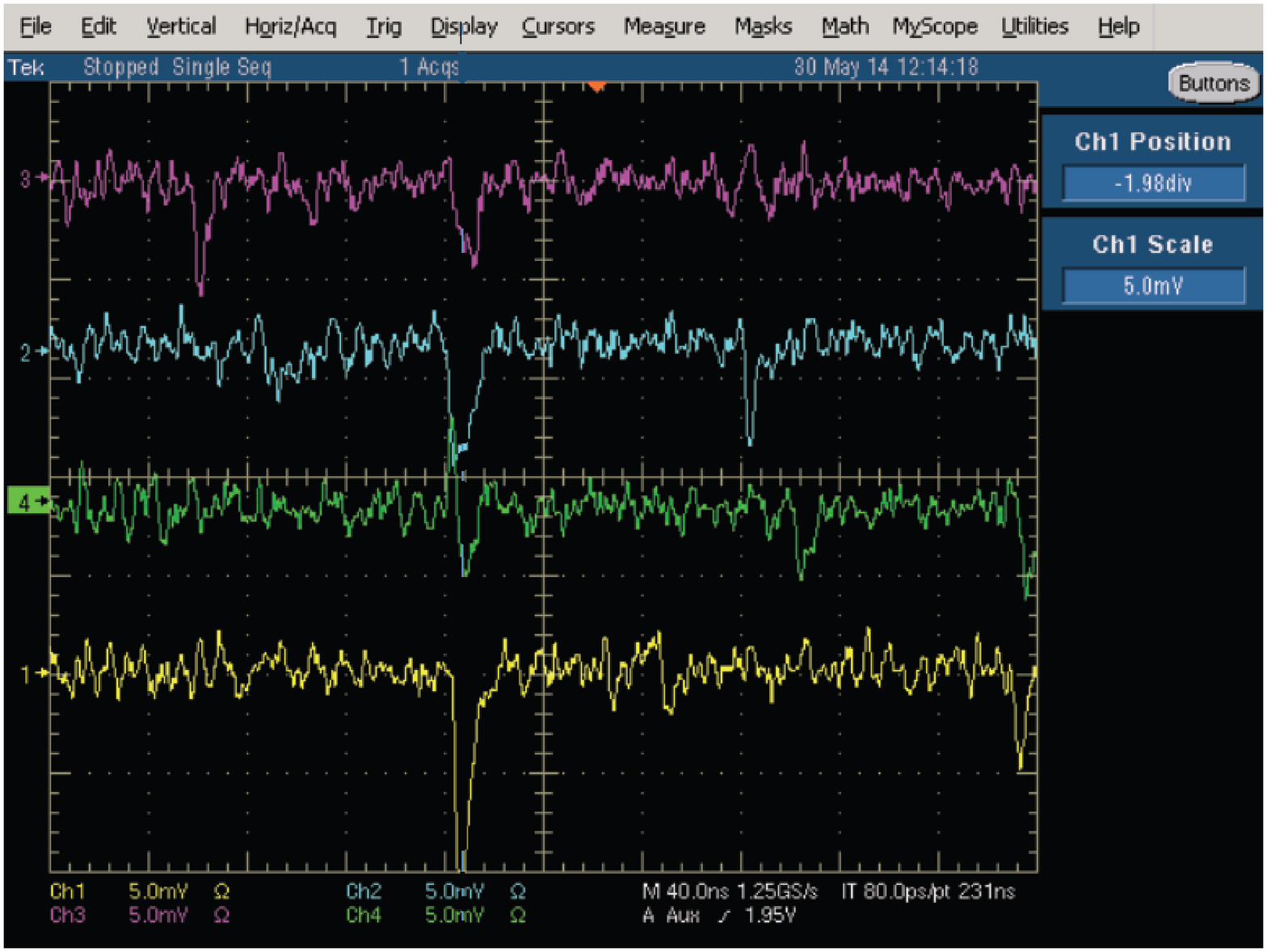}
\end{tabular}
\caption{Screenshots from the oscilloscope's monitor.
{\it Left panel}: Signals sampled in the dark;
{\it Right panel}: Signals sampled in
coincidence with the laser pulse.}
\label{Fig_4}
\end{figure}

We tested the coupling of the SiPM matrix with the electronics
by means of a pulsed-mode laser light.
Four out of the 16 signals were sent to a TDS5104B oscilloscope,
with a signal sampling of 5 GHz.
Two screen-shots are shown in Figure \ref{Fig_4}.
The {\it left panel} refers to the dark count
signals. Here, the average amplitude per single
p.e. is about 5 mV and the average distance time
is about 200 ns, corresponding to a total
dark rate of about 500 kHz/mm$^2$. It is worth
noting that the SiPMs adopted for this analysis
represent a very preliminary production by FBK.
More recently developed sensors show an
improvement in the dark count rate
by an order of magnitude.
The {\it right panel} of Figure \ref{Fig_4}
shows the signal sampling in the case of a laser event,
characterized by the four SiPM signals being
coincident in time.

\begin{figure}[t!]
\centering
\begin{tabular}{cc}
\includegraphics[height=0.3\textwidth,bb=0 0 590 490,clip]{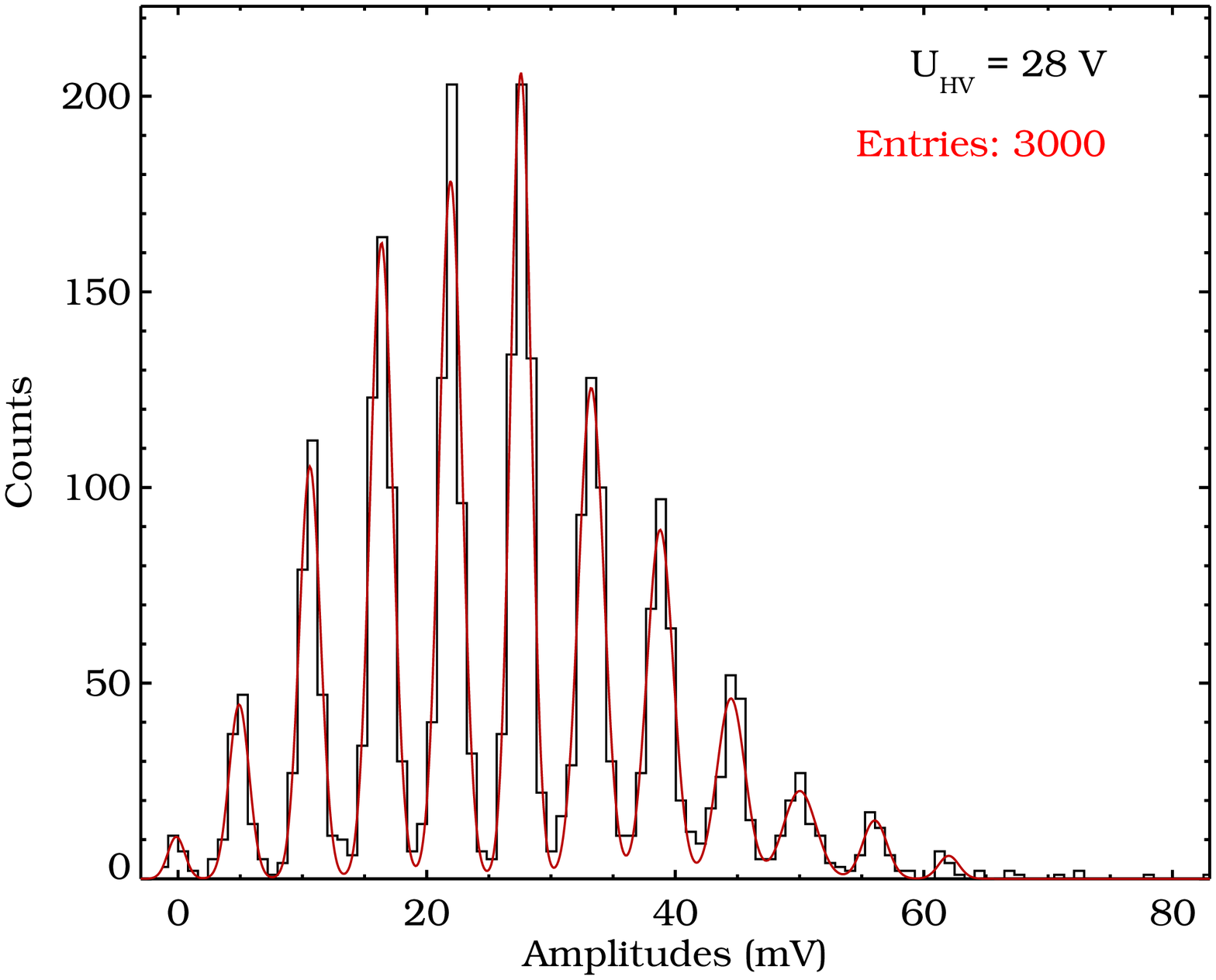} &
\includegraphics[height=0.3\textwidth,bb=0 0 590 490,clip]{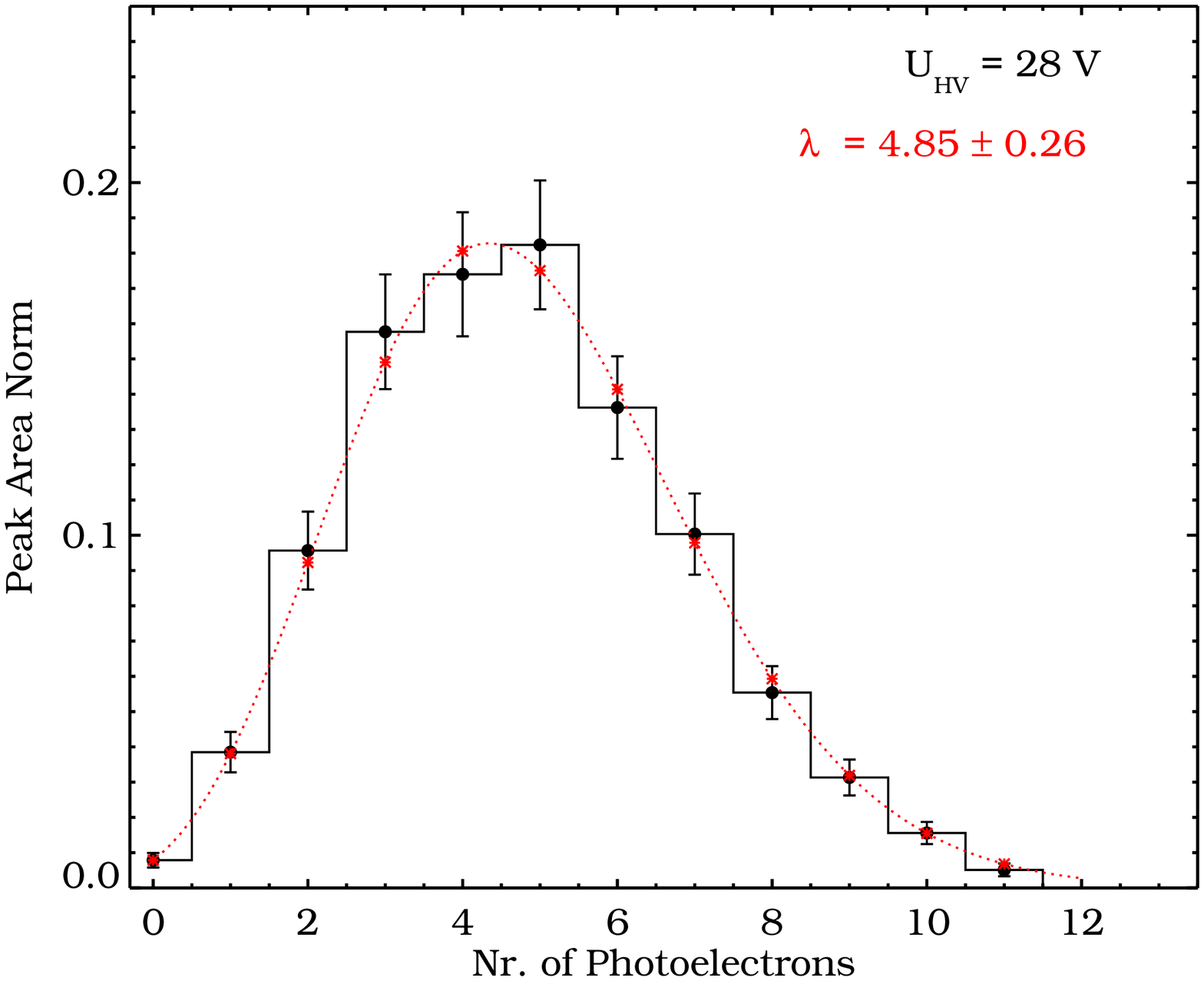} \\
\includegraphics[height=0.3\textwidth,bb=0 0 590 490,clip]{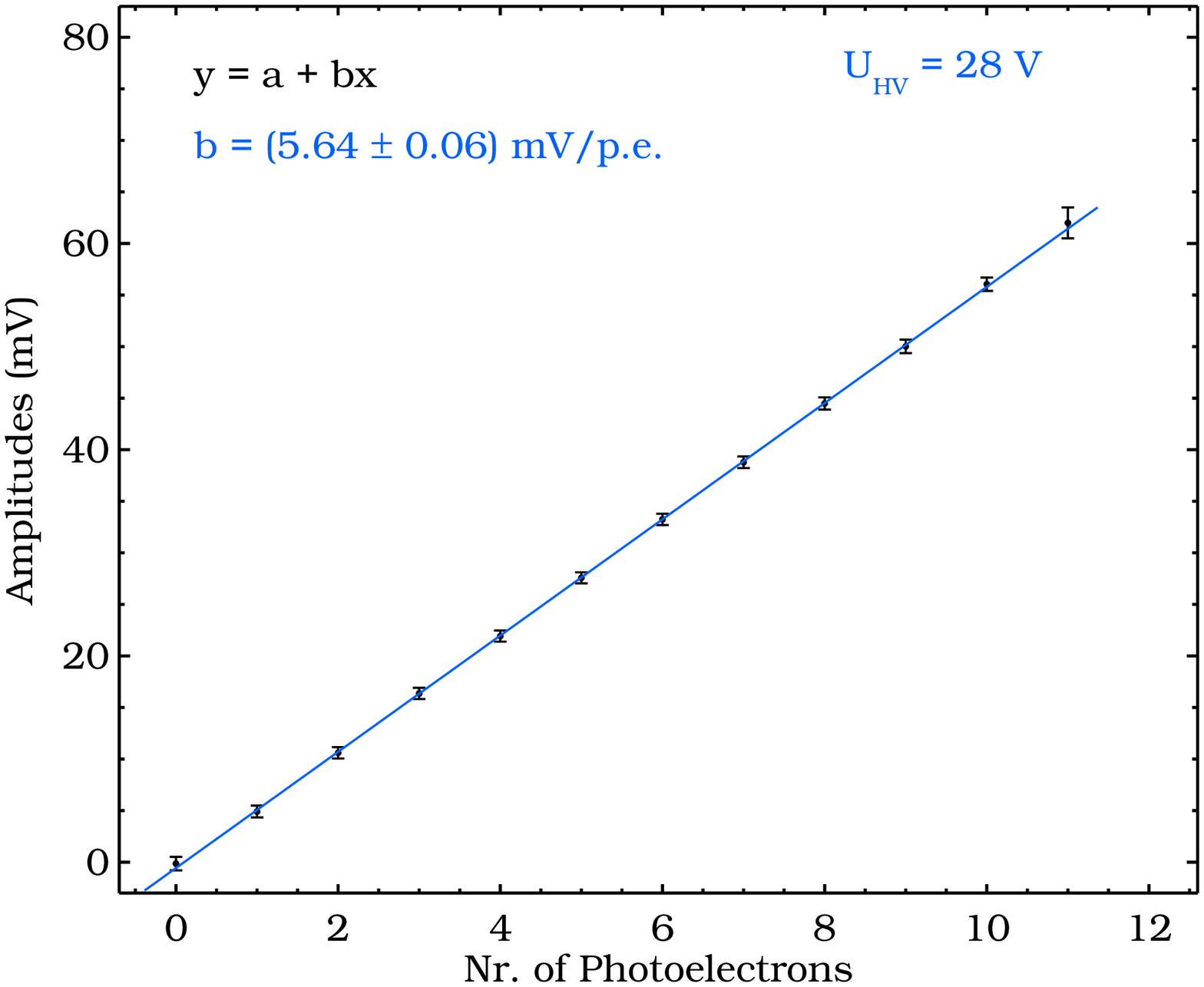} &
\includegraphics[height=0.3\textwidth,bb=0 0 590 490,clip]{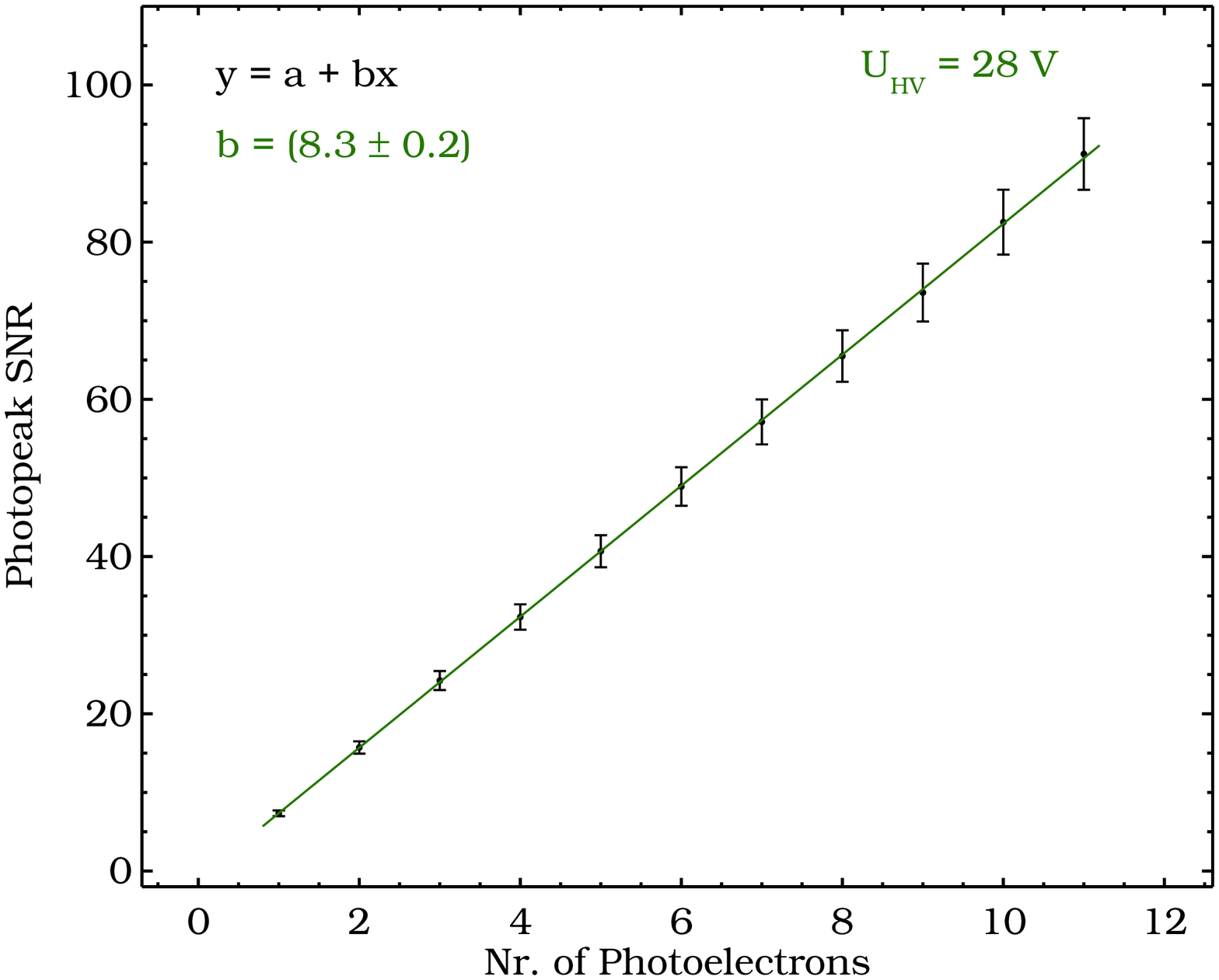}
\end{tabular}
\caption{
{\it Top left panel:} Amplitude spectrum from a single SiPM.
{\it Top right panel:} Amplitude distribution.
{\it Bottom left panel:} Gain.
{\it Bottom right panel:} SNR.
}
\label{Fig_5}
\end{figure}

We studied the amplitude of the signal from a single SiPM
by covering all other SiPMs.
An example amplitude spectrum is shown in the {\it top left panel}
of Figure \ref{Fig_5}. The amplitude distribution
of all p.e. events clearly exhibits
a Poissonian shape, as can be seen {\it top right panel}
of Figure \ref{Fig_5}.
We also studied the linearity of the SiPM electronics chain.
The {\it bottom left panel} of Figure \ref{Fig_5} displays
a linear fit, which indicates a good linear correlation
of the signal amplitude vs. the number of converted photo-electrons.
Finally we evaluated the signal-to-noise ratio (SNR) by
calculating the ratio between the gain (i.e. the position
of the single p.e. amplitude)
to the width of the electronic noise (see the
{\it bottom right panel} of Figure \ref{Fig_5}).

In summary, we conclude that our device shows a gain
of 5.6 mV/p.e. with a SNR of about 8.3 for an OV of roughly 2 V.
\section{Future Work}
\begin{figure}[ht!]
\centering
\begin{tabular}{cc}
\includegraphics[height=0.29\textwidth,bb=0 50 1040 550,clip]{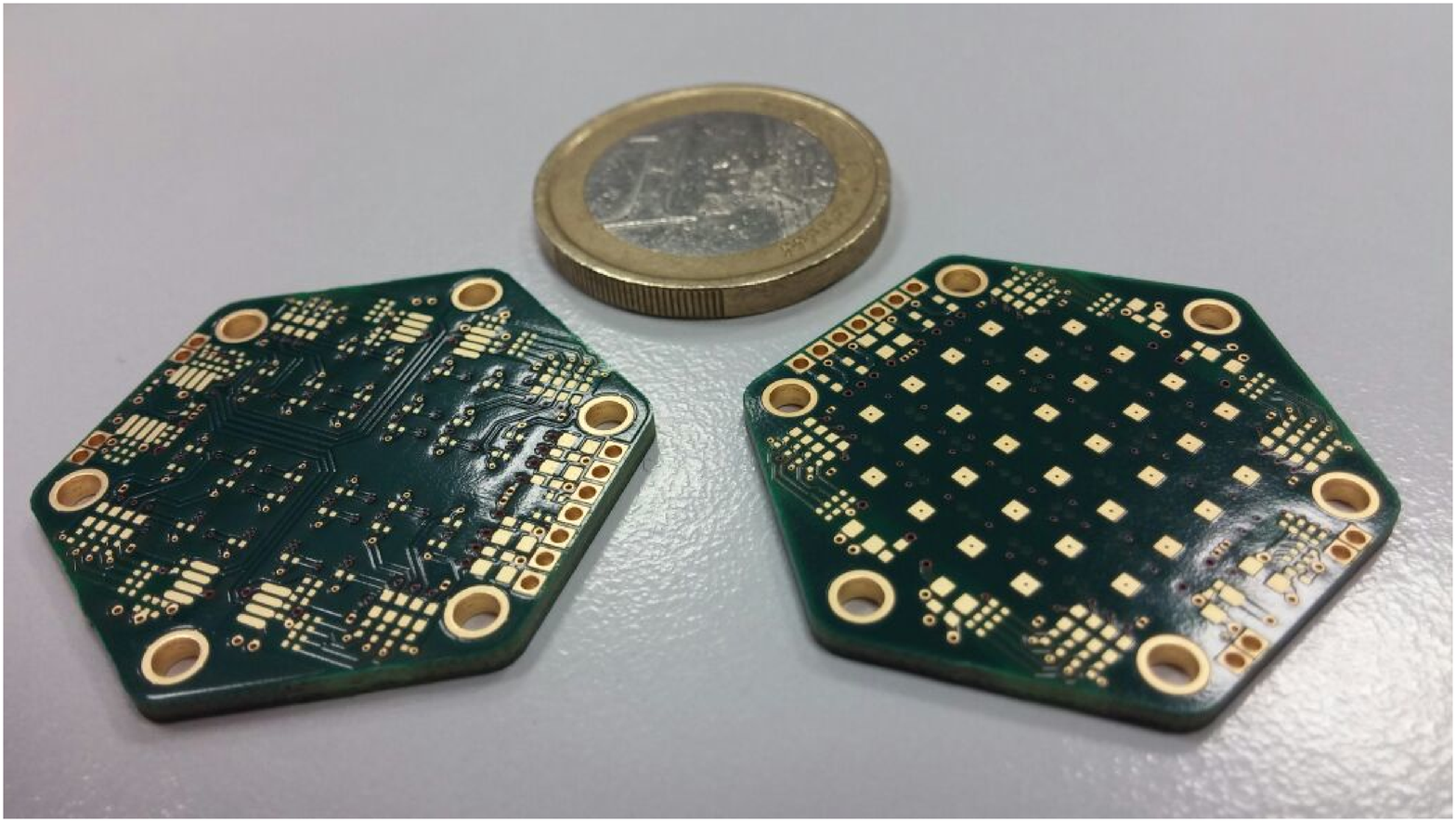} &
\includegraphics[height=0.29\textwidth,bb=150 0 840 585,clip]{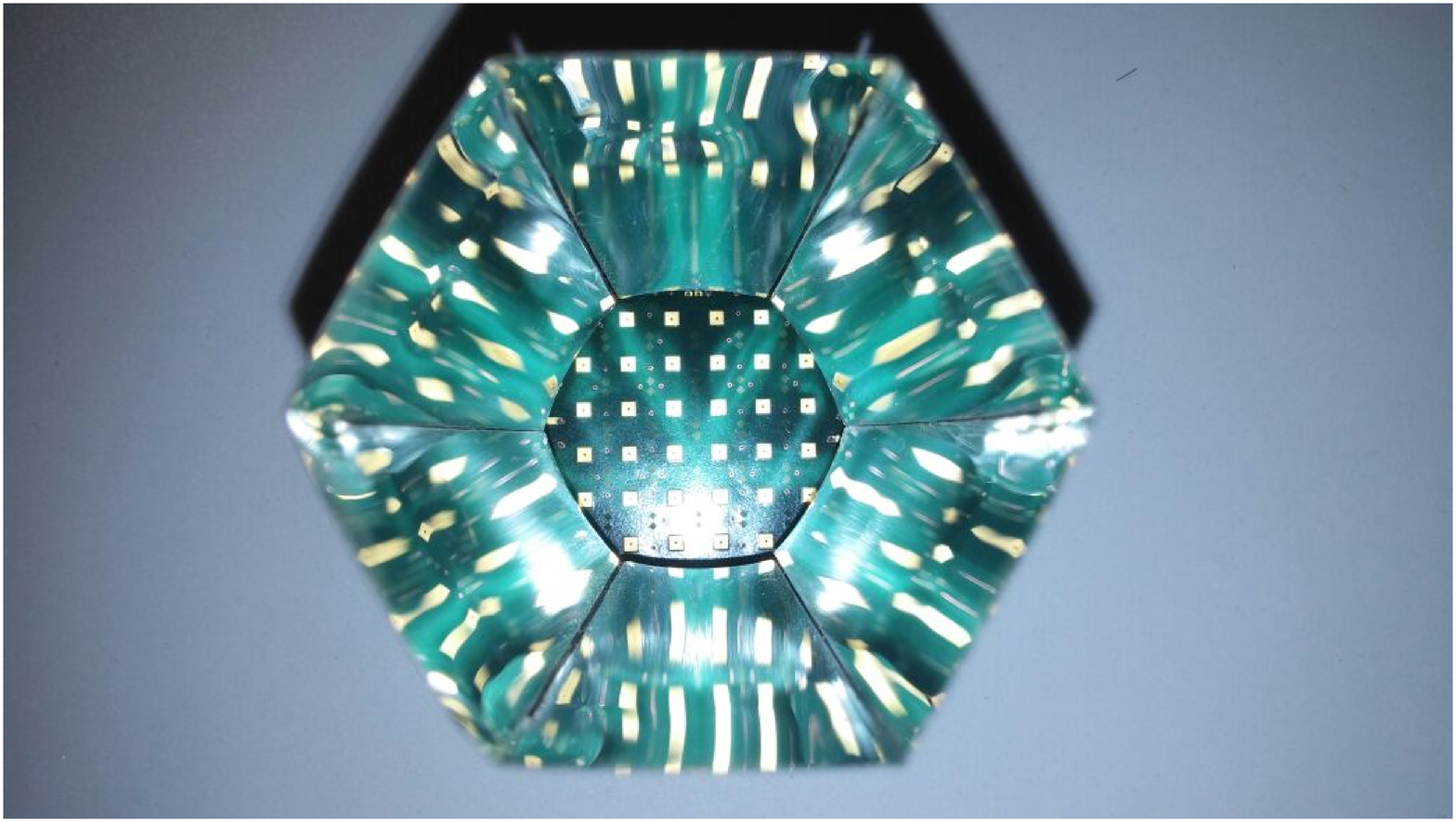} 
\end{tabular}
\caption{High dimensions pixel:
{\it Left panel:} array of SiPMs and electronics;
{\it Right panel:} the SiPM array seen through the light guide.}
\label{Fig_Last}
\end{figure}

INFN is currently producing a cluster of 
7 or more 3 $\times$ 3 (or  6 $\times$ 6) 
SiPMs of 1 inch diameter, which can be
suitable for telescopes with big cameras, like 
CTA's large size telescopes (LSTs).
The {\it left panel} of
Figure \ref{Fig_Last} shows a preliminary 
design of a PCB for housing a matrix of 16 
SiPMs and an electronics optimized
to collect the signals of all these 
sensors and to fed out the signal 
of only one pixel. The coupling 
of this high dimensions pixel with a 
light guide (shown in the {\it right panel}) 
is also being exploited.
The first prototypes have been produced 
with a cooperation between INFN Padova and
INFN Perugia and will be tested on the MAGIC
telescope.

Moreover, FBK is continuously producing and upgrading
devices with a reduced dark count rate,
an increased PDE and an optimized geometry,
in order to reduce the dead area zone.
Further studies are in progress
in order to optimize the current signal tail
in order to limit the impact
of the diffuse light background
to the trigger efficiency and signal
reconstruction. For this purpose, more tests
are currently under investigation.

\end{document}